%% file: main.tex
\documentclass[a4paper,twoside]{article}

\usepackage{epsfig}
\usepackage{subcaption}
\usepackage{calc}
\usepackage{amssymb}
\usepackage{amstext}
\usepackage{amsmath}
\usepackage{amsfonts}
\usepackage{amsthm}
\usepackage{multicol}
\usepackage{pslatex}
\usepackage{apalike}

\usepackage[scaled=0.8]{helvet}    
\usepackage[misc,geometry]{ifsym} 
\usepackage{graphicx}
\usepackage[dvipsnames]{xcolor}
\usepackage{tikz}
\usetikzlibrary{trees,snakes,arrows}
\usetikzlibrary{shapes,chains}
\usetikzlibrary{positioning}
\usepackage{hyperref}
\hypersetup{
    colorlinks=true,
    linkcolor=blue,
    citecolor=blue,
    urlcolor=blue
}
\usepackage[nospace]{cite}

\usepackage{SCITEPRESS}     

\input{macros}
\begin{document}
\title{Formal Analysis of EDHOC Key Establishment for Constrained IoT Devices}
\author{
    \authorname{
        Karl Norrman\sup{1}\sup{,}\sup{2}\orcidAuthor{0000-0003-0164-1478}
        , Vaishnavi Sundararajan\sup{3} and
        Alessandro Bruni\sup{4}
    }
\affiliation{\sup{1}KTH Royal Institute of Technology, Stockholm, Sweden}
\affiliation{\sup{2}Ericsson Research, Security, Stockholm, Sweden}
\affiliation{\sup{3}University of California Santa Cruz, USA}
\affiliation{\sup{4}IT University of Copenhagen, Copenhagen, Denmark}
    \email{karl.norrman@ericsson.com, vasundar@ucsc.edu, brun@itu.dk}
}


\abstract{
Constrained IoT devices are becoming ubiquitous in society
and there is a need for secure communication protocols that respect the
constraints under which these devices operate.
\mEdhoc{} is an authenticated key establishment protocol for constrained IoT
devices, currently being standardized by the Internet Engineering Task
Force (IETF).
A rudimentary version of \mEdhoc{} with only two key establishment methods was
formally analyzed in 2018.
Since then, the protocol has evolved significantly and several new key
establishment methods have been added.
In this paper, we present a formal analysis of all \mEdhoc{} methods in an
enhanced symbolic Dolev-Yao model using the \mTamarin{} tool.
We show that not all methods satisfy the authentication notion
injective of agreement, but that they all do satisfy a notion of implicit
authentication, as well as Perfect Forward Secrecy (PFS) of the session key material.
We identify other weaknesses to which we propose improvements.
For example, a party may intend to establish a session key with a certain peer,
but end up establishing it with another, trusted but compromised, peer.
We communicated our findings and proposals to the IETF, which
has incorporated some of these in newer versions of the standard.
}

\onecolumn \maketitle \normalsize \setcounter{footnote}{0} \vfill
\section{\uppercase{Introduction}}
\label{sec:introduction}

As IoT devices become more prevalent and get involved in progressively sensitive
functions in society, the need to secure their communications
becomes increasingly important.
Most security analysis has focused on computationally
strong devices, such as cars and web-cameras, where existing protocols like
\mDandTls{} suffice.
Constrained devices, on the other hand, which operate under severe
bandwidth and energy consumption restrictions, have received much less
attention.
These devices may be simple sensors, which only relay environment
measurements to a server every hour, but need to function autonomously without
maintenance for long periods of time.
The IETF standardized the Object Security for
Constrained RESTful Environments (\mOscore{}) protocol to secure communications
between constrained devices~\cite{rfc8613}.
However, the \mOscore{} protocol requires a pre-established security context.
The IETF has been discussing requirements and mechanisms for a key
exchange protocol, named Ephemeral Diffie-Hellman Over COSE (\mEdhoc), for
establishing \mOscore{} security contexts.
Naturally, \mEdhoc{} must work under the same constrained requirements as
\mOscore{} itself.
While not all use cases for \mEdhoc{} are firmly set, the overall goal is to
establish an \mOscore{} security context, under message size limitations.
It is therefore important to ensure that \mEdhoc{} satisfies fundamental
security properties expected from a key exchange protocol.
%

The first incarnation of \mEdhoc{} appeared in March 2016.
It contained two different key establishment methods, one based on a
pre-shared Diffie-Hellman (DH) cryptographic core%
\footnote{By a \emph{cryptographic core}, or simply core, we mean an academic protocol,
without encodings or application specific details required by an industrial
protocol.
A \emph{key establishment method} is a core with some such details added.
}
and a second based on a
variant of challenge-response signatures in the style of
\mOptls{}~\cite{DBLP:conf/eurosp/KrawczykW16}.
\mEdhoc{} is therefore a framework of several key establishment methods.
In May 2018, the core based on challenge-response signatures was replaced by
one based on \mSigma{} (SIGn-and-MAc)~\cite{sigma,bruni-analysis-selander-ace-cose-ecdhe-08}.
Since then the protocol has undergone significant changes.
Three new cores, mixing challenge-response signatures and regular signatures for
authentication, were added~\cite{our-analysis-selander-lake-edhoc-00}.

We formulate and formalize a security model covering all four key
establishment methods, which is important especially since the
\mSpec{}~\cite{our-analysis-selander-lake-edhoc-00} lacks a clear description
of the intended security model and overall security goals.

We perform the analysis in a symbolic Dolev-Yao model.
In this framework, we model messages as terms in an algebra, with operations such
as encryption modelled as functions on these terms.
These functions are assumed perfect, e.g., one cannot decrypt an encrypted
message without access to the key.
The adversary, while unable to break encryption or reverse hashing, is modelled
as the network.
That is, the adversary, can block reroute, replay and modify messages at will.
A symbolic model like this, while slightly severe an abstraction, still allows
us to analyze \mEdhoc{} for logical flaws without incurring the complexity of
a computational model.
The standardization process is ongoing, with the authors releasing newer
versions of the \mSpec{} (see Section~\ref{sec:newdrafts} for more detail
about how these versions differ from the one analyzed here).
%

\subsection{Contributions}
\label{sec:contributions}
In this paper, we formally analyze the \mEdhoc{} protocol (with its four key
establishment methods) using the \mTamarin{} tool~\cite{DBLP:conf/cav/MeierSCB13}.
We present a formal model we constructed of the protocol as given in the
\mSpec{}~\cite{our-analysis-selander-lake-edhoc-00}.

We give an explicit adversary model for the protocol and verify
properties such as session key material and entity authentication, and perfect
forward secrecy, for all four methods.

The model itself is valuable as a basis for verifying further updates in the
ongoing standardization.
It is publicly available~\cite{edhocTamarinRepo}.
It took several person-months to interpret the
specification and construct the model.
Termination requires a hand-crafted proof oracle to guide \mTamarin{}.

We show that not all \mEdhoc{}'s key establishment
methods provide authentication according to the injective agreement definition
on the session key material, and none on the initiator's identity.
However, we show that all methods fulfill an implicit agreement property
covering the session key material and the initiator's identity.
We identify a number of subtleties, ambiguities and weaknesses in the
specification.
For example, the authentication policy requirements allow situations where a
party establishes session key material with a trusted but compromised peer, even
though the intention was to establish it with a different trusted party.
We provide remedies for the identified issues and have
communicated these to the IETF and the specification authors, who have
incorporated some of our suggestions and are currently considering how to deal
with the remaining ones.
%

\subsection{Comparison with Related Work}
The May 2018 version of \mEdhoc{} was formally analyzed by
\cite{DBLP:conf/secsr/BruniJPS18} using the \mProverif{}
tool~\cite{DBLP:conf/csfw/Blanchet01}.
Their analysis covered a pre-shared key authenticated core and one
based on \mSigma.
The properties checked for therein were secrecy, PFS and integrity of
application data, identity protection against an active adversary,
and strong authentication.

In contrast to the key establishment methods analyzed by Bruni et~al., which
were based on the well-understood pre-shared key DH and \mSigma{} protocols,
the three newly added
methods combine two unilateral authentication protocols with the goal to
constructing mutual authentication protocols.
Combining two protocols, which individually provide unilateral authentication,
is not guaranteed to result in a secure mutual authentication
protocol~\cite{DBLP:conf/ccs/Krawczyk16}.
Consequently, even though the framework is similar to the one analyzed by Bruni
et~al., the cryptographic underpinnings have significantly increased in
complexity, and is using mechanisms which have not previously been formally analyzed.
The set of properties we check for is also different.
Our analysis is further carried out using a different tool,
namely \mTamarin; different kinds of strategies to formulate and
successfully analyze the protocol are required when working with this tool.
%

\section{\uppercase{The \mEdhoc{} Protocol}}
\label{sec:edhoc}
\input{edhocProtocol}

\section{\uppercase{Formalization and Results}}
\label{sec:formalization}
\input{edhocFormalization}

\section{\uppercase{Discussion}}
\label{sec:discussion}
There are a few places where \mEdhoc{} can be improved,
which we found during this work and communicated to the authors.
We discuss them below.
%

\subsection{Unclear Intended Use}
\label{sec:unclearProtocolUse}
The \mEdhoc{} \mSpec{} lists several security goals, but they are
imprecise and difficult to interpret due to lack of context and intended usage
descriptions.
Without knowing how the protocol is to be used,
it is not clear whether the listed security goals are the most important ones
for constrained IoT devices.

The abstract goal of \mEdhoc{} is simple: establish an \mOscore{} security
context using few roundtrips and small messages.
From that, the design of \mEdhoc{} is mainly driven by what
can be achieved given the technical restrictions.
Focusing too much on what can be achieved within given restrictions, and paying
too little attention to the use cases where the
protocol is to be used and their specific goals, risks resulting in
sub-optimal trade-offs and design decisions.

\mEdhoc{} is intended to cover a variety of use cases, many of which are
difficult to predict today.
However, this does not
prevent collecting \emph{typical} use cases and user stories
to identify more specific security goals that will be important in most cases.

While constructing our model, we made up simple user stories to identify
security properties of interest.
Several of these revealed subtleties and undefined aspects of \mEdhoc{}.
We informed the \mEdhoc{} authors, who addressed these aspects in the
\mSpec{}.

\subsubsection{(Non-)Repudiation}
An access control solution for a nuclear power-plant may need to log who is
passing through a door, whereas it may be undesirable for, say, a coffee
machine to log a list of users along with their coffee preferences.
Via this simple thought experiment, we realized that the \mSpec{} did not
consider the concept of (non)-repudiation.
In response, the authors of the \mSpec{} added a paragraph discussing how
different methods relate to (non)-repudiation.

\subsubsection{Unintended Peer Authentication}
According to Section~3.2 of the \mSpec{}, parties must be configured
with a policy restricting the set of peers they run \mEdhoc{} with.
However, the initiator is not required to verify that the \mIdcredr{} received
in the second message is the same as the one intended.
The following attack scenario is therefore possible.

Suppose someone has configured all devices in their home to be in the
allowed set of devices, but that one of the devices ($A$) is compromised.
If another device $B$, initiates a connection to a third device $C$, the
compromised device $A$ may interfere by responding in $C$'s place, blocking
the legitimate response from $C$.
Since $B$ does not verify that the identity indicated in the second message
matches the intended identity $C$, and device $A$ is part of the allowed set,
$B$ will complete and accept the \mEdhoc{} run with device $A$ instead of the
intended $C$.
The obvious solution is for the initiator to match \mIdcredr{} to the intended
identity indicated by the application, which we included in our model.
We have communicated this to the \mEdhoc{} authors and they are considering
a resolution.
%

\subsection{Unclear Security Model}
We argue that the \mSpec{} gives too little information about what capabilities
an adversary is assumed to have, and that this leads to unclear design goals and
potentially sub-optimal design.

Even though \mEdhoc{} incorporates cryptographic cores from different academic
security protocols, its design does not take into account the adversary models
for which these protocols were designed.
For example, \mOptls{}, whose cryptographic core is essentially the same
as the \mStat{} authentication method, is designed to be secure in the CK
model~\cite{DBLP:conf/crypto/CanettiK02}.
The CK security model explicitly separates the secure storage of long-term
keys from storage of session state and ephemeral keys.
This is appropriate for modelling the use of secure modules.

The \mEdhoc{} authors indicated to us that it was
not necessary to consider compromised ephemeral keys separately from
compromised long-term keys.
The rationale is that \mSigma{} cannot protect against compromised ephemeral
keys~\cite{personalCommunication}.
That rationale is presumably based on the fact that the \mSigSig{} method is
closely modeled on the \mSigmaI{} variant of \mSigma{}, and that it would be
preferable to obtain a homogeneous security level among the \mEdhoc{}
methods.
That is only true, however, when restricting attention to session key
confidentiality of an ongoing session.
Secure modules provide value in other ways, for example, by allowing
constructions with Post-Compromise Security (PCS) guarantees.
We discussed this with the authors, and
the latest version of the \mSpec{}~\cite{latest-ietf-lake-edhoc-05} includes
recommendations on storage of long-term keys and operations on these inside a
secure module.
%

\subsection{Session Key Material}
\label{sec:sessionKeyMaterial}
\mEdhoc{} establishes session key material, from which session keys
can be derived using the \mEdhoc{}-Exporter.
The session key material is affected by \mGxy{}, and if a party uses the
\mStat{} authentication method, also by that party's secret static long-term key.
As shown in Section~\ref{sec:formalization}, mutual injective agreement cannot
be achieved for $P_I$.
If this property is not important for constrained IoT devices which cannot use
any of the other methods, then one can simply accept that the methods have
different authentication strengths.
Otherwise, this is a problem.

We identified three alternatives for resolving this.
One alternative is to include \mIdcredi{}, or its hash, in the first and
second messages.
This would, however, increase message sizes and prevent initiator identity
protection, which are grave concerns for \mEdhoc{}.
A second alternative is to not derive the session key material from $P_I$.
Doing so, however, deviates from the design of \mOptls{} (and similar protocols
from which the \mStat{}-based methods are derived), where the inclusion of
$P_I$ plays a crucial part in the security proof of resistance against
initiator ephemeral key compromise.
The third alternative is to include a fourth message from responder to initiator,
carrying a MAC based on a key derived from session key material including $P_I$.
Successful MAC verification guarantees
to the initiator that the responder injectively agrees on $P_I$.
We presented the options to IETF, and they decided to add a
fourth message as an option in the latest version of the
\mSpec{}~\cite{latest-ietf-lake-edhoc-05}.

We verified that all methods
enjoy a common, but weaker, property: mutual implicit agreement
on all of $P_e, P_I$ and $P_R$, where applicable.
%

\section{\uppercase{Conclusions and Future Work}}
\label{sec:conclusions}
\label{sec:newdrafts}
We formally modeled all four
methods of the \mEdhoc{} \mSpec{} using \mTamarin.
We formulated several important security properties and identified precise
adversary models in which we verified these.
The properties are shown in Table~\ref{tab:props}.
Mutual injective agreement covers the set of parameters $S_P$:
responder identity, roles, session key material (except for $P_I$ when
initiator uses the \mStat{} authentication
method), context identifiers \mCi{} and \mCr, and cipher suites \mSuites.
The responder in addition is ensured agreement on the initiators identity and
$P_I$, i.e., on the set $S_F$.
Implicit agreement covers all previously mentioned parameters for both peers.
Verification of all lemmas, including model validation lemmas, took 42 minutes
on an Intel Core i7-6500U 2.5GHz using two cores.
Mutual entity authentication, UKS- and KCI resistance can be inferred
from the verified properties.
\begin{table*}[h!]
        \centering
        \caption{Verified properties. $S_P$ contains
            roles, responder identity, session key material (excluding
            $P_I$), \mCi, \mCr, and \mSuites. $S_F$ is $S_{P}$,
            the initiator identity, and $P_I$.}
        \label{tab:props}
        \begin{tabular}{|l|c|c|c|c|}
                \hline
                & \mSigSig & \mSigStat & \mStatSig & \mStatStat \\
                \hline
                Injective agreement for I & $S_F$ & $S_F$ & $S_P$ & $S_P$\\
                Injective agreement for R & $S_F$ & $S_F$ & $S_F$ & $S_F$\\
                Implicit agreement for I & $S_F$ & $S_F$ & $S_F$ & $S_F$\\
                Implicit agreement for R & $S_F$ & $S_F$ & $S_F$ & $S_F$\\
                PFS for session key material & \cm & \cm & \cm & \cm\\
                \hline
        \end{tabular}
\end{table*}

Further, we identified a situation where initiators may establish an \mOscore{}
security context with a different party than the application using \mEdhoc{}
intended, and proposed a simple mitigation.
We discussed how the IETF may extract and better define security properties to
enable easier verification.

We verified each method in isolation.
Verifying security under composition is left as future work.

In this work, we have analyzed the \mEdhoc{} version as of July
2020~\cite{our-analysis-selander-lake-edhoc-00}.
There are newer versions, with the most recent version as
of February 2021~\cite{latest-ietf-lake-edhoc-05}.
However, the changes to the protocol over these versions are not
particularly significant for our analysis.
%

\section*{ACKNOWLEDGEMENTS}
This work was partially supported by
the Wallenberg AI, Autonomous Systems and Software Program (WASP) funded by
the Knut and Alice Wallenberg Foundation.
We are grateful to G\"oran Selander, John Mattsson and Francesca Palombini for
clarifications regarding the specification.
%

\bibliographystyle{apalike}
{\small
    \bibliography{refComp}
}

\end{document}

%% file: macros.tex
\usepackage{ifthen}

\usepackage{todonotes}

\definecolor{cbnavy}{RGB}{15, 32, 128}
\definecolor{cborange}{RGB}{245, 121, 58}
\definecolor{cbsky}{RGB}{133, 192, 249}

\newcommand{\mFunStyle}[1]{\textsf{#1}}
\newcommand{\mConstStyle}[1]{\textsf{#1}}

\newcommand{\mMethodStyle}[1]{\mConstStyle{#1}}
\newcommand{\mProtocolStyle}[1]{\text{#1}}

\newcommand{\mSpec}{specification}  

\newcommand{\mRevLTK}{\ensuremath{\mathbf{A}_\mFunStyle{LTK}}}

\newcommand{\mRevEph}{\ensuremath{\mathbf{A}_\mFunStyle{Eph}}}
\newcommand{\mIStart}{\ensuremath{\mathbf{I}_\mFunStyle{S}}}
\newcommand{\mIComplete}{\ensuremath{\mathbf{I}_\mFunStyle{C}}}
\newcommand{\mRStart}{\ensuremath{\mathbf{R}_\mFunStyle{S}}}
\newcommand{\mRComplete}{\ensuremath{\mathbf{R}_\mFunStyle{C}}}
\newcommand{\mPredPfs}{\ensuremath{\mathbf{PFS}}}
\newcommand{\mPredInjI}{\ensuremath{\mathbf{InjAgree}_I}}
\newcommand{\mPredInjR}{\ensuremath{\mathbf{InjAgree}_R}}
\newcommand{\mPredImpI}{\ensuremath{\mathbf{ImpAgree}_I}}
\newcommand{\mPredImpR}{\ensuremath{\mathbf{ImpAgree}_R}}
\newcommand{\mK}{\ensuremath{\mathcal{K}}}
\DeclareMathOperator{\mLogicDot}{.}

\newcommand{\mTamarin}{\mProtocolStyle{Tamarin}}
\newcommand{\mProverif}{\mProtocolStyle{ProVerif}}
\newcommand{\mEdhoc}{\mProtocolStyle{EDHOC}}
\newcommand{\mOscore}{\mProtocolStyle{OSCORE}}
\newcommand{\mSigma}{\mProtocolStyle{SIGMA}}
\newcommand{\mSigmaI}{\mProtocolStyle{SIGMA\nobreakdash-I}}

\newcommand{\mCose}{\mProtocolStyle{COSE}}

\newcommand{\mHkdf}{\mProtocolStyle{HKDF}}
\newcommand{\mHkdfExtract}{\mProtocolStyle{HKDF\nobreakdash-extract}}
\newcommand{\mHkdfExpand}{\mProtocolStyle{HKDF\nobreakdash-expand}}

\newcommand{\mAead}{\mProtocolStyle{AEAD}}

\newcommand{\mOptls}{\mProtocolStyle{OPTLS}}

\newcommand{\mDandTls}{\mProtocolStyle{(D)TLS}}

\newcommand{\mStat}{\mMethodStyle{STAT}}
\newcommand{\mSig}{\mMethodStyle{SIG}}

\newcommand{\mStatStat}{\mMethodStyle{STAT-STAT}}
\newcommand{\mStatSig}{\mMethodStyle{STAT-SIG}}
\newcommand{\mSigStat}{\mMethodStyle{SIG-STAT}}
\newcommand{\mSigSig}{\mMethodStyle{SIG-SIG}}

\newcommand{\mXor}{\mConstStyle{XOR}}
\newcommand{\mSuites}{\ensuremath{S_I}}
\newcommand{\mMethod}{\ensuremath{M}}
\newcommand{\mCi}{\ensuremath{C_I}}
\newcommand{\mCr}{\ensuremath{C_R}}
\newcommand{\mGi}{\ensuremath{Q_{s,I}}}
\newcommand{\mGr}{\ensuremath{Q_{s,R}}}
\newcommand{\mPriv}[1]{\ensuremath{d_{s,#1}}}
\newcommand{\mPub}[1]{\ensuremath{Q_{s,#1}}}
\newcommand{\mPrivE}[1]{\ensuremath{d_{e,#1}}}
\newcommand{\mPubE}[1]{\ensuremath{Q_{e,#1}}}
\newcommand{\mX}{\ensuremath{d_{e,I}}}
\newcommand{\mY}{\ensuremath{d_{e,R}}}

\newcommand{\mGx}{\ensuremath{Q_{e,I}}}
\newcommand{\mGy}{\ensuremath{Q_{e,R}}}
\newcommand{\mGxy}{\ensuremath{P_e}}
\newcommand{\mSessKey}{\ensuremath{Z}}

\newcommand{\mSign}[1]{\ensuremath{\mathit{sign_{#1}}}}
\newcommand{\mVf}[1]{\ensuremath{\mathit{vf_{#1}}}}
\newcommand{\mDH}{\ensuremath{\mathit{dh}}}

\newcommand{\mTHtwo}{\ensuremath{\mathit{th}_2}}
\newcommand{\mKtwoe}{\ensuremath{K_\mathit{2e}}}
\newcommand{\mKtwom}{\ensuremath{K_\mathit{2m}}}

\newcommand{\mKthreeae}{\ensuremath{K_\mathit{3ae}}}
\newcommand{\mKthreem}{\ensuremath{K_\mathit{3m}}}

\newcommand{\mTHthree}{\ensuremath{\mathit{th}_3}}

\newcommand{\mCredi}{\ensuremath{Q_{s,I}}}
\newcommand{\mCredr}{\ensuremath{Q_{s,R}}}
\newcommand{\mHash}{\ensuremath{h}}

\newcommand{\mTHfour}{\ensuremath{\mathit{th}_4}}
\newcommand{\mAuthi}{\ensuremath{\mathit{Auth}_I}}
\newcommand{\mAuthr}{\ensuremath{\mathit{Auth}_R}}

\newcommand{\mMactwo}{\ensuremath{\mathit{MAC}_2}}
\newcommand{\mMacthree}{\ensuremath{\mathit{MAC}_3}}

\newcommand{\mSigthree}{\ensuremath{\mathit{Sig}_3}}

\newcommand{\mMsgone}{\ensuremath{m_1}}
\newcommand{\mMsgtwo}{\ensuremath{m_2}}
\newcommand{\mMsgthree}{\ensuremath{m_3}}

\newcommand{\mCipher}{\mConstStyle{CIPHERTEXT\_2}}

\newcommand{\mADone}{\ensuremath{\mathit{ad}_1}}
\newcommand{\mADtwo}{\ensuremath{\mathit{ad}_2}}
\newcommand{\mADthree}{\ensuremath{\mathit{ad}_3}}

\newcommand{\mPRK}{\ensuremath{\mathit{PRK}}}
\newcommand{\mPRKtwo}{\ensuremath{\mPRK_\mathit{2e}}}
\newcommand{\mPRKthree}{\ensuremath{\mPRK_\mathit{3e2m}}}
\newcommand{\mPRKfour}{\ensuremath{\mPRK_\mathit{4x3m}}}

\newcommand{\mIdcredi}{\ensuremath{ID_I}}
\newcommand{\mIdcredr}{\ensuremath{ID_R}}

\newcommand{\cm}{\checkmark}

\newcommand{\msg}[4]{\draw[->,thick] ([yshift=-#1]#2.south) coordinate (l1)--(l1-|#3) node[midway, above]{#4}}
\newcommand{\action}[3]{\node[draw,thick,fill=white,align=center,below={#1} of {#2}]{#3}}

\newcommand{\ifarrow}[1][]{\ifthenelse{\equal{#1}{}}{\rightarrow}{-\hspace{-3.8pt}[{#1}]\hspace{-4.3pt}\rightarrow}}
\newcommand{\semarrow}[1][]{\ifthenelse{\equal{#1}{}}{\Rightarrow}{=\hspace{-3.0pt}[{#1}]\hspace{-3pt}\Rightarrow}}

\newcommand{\mT}[1]{\text{\texttt{#1}}}

%% file: edhocProtocol.tex
We now present the structure of the protocol using notation for key material,
elliptic curve operations and
identities mostly adopted from NIST SP 800-56A Rev. 3~\cite{sp800-56a-rev3}.
One notable difference is that we refer to the two roles executing the protocol
as the initiator $I$ and the responder $R$.
We do this to avoid confusing roles with the parties taking them on.
Values in the analysis are subscripted with $I$ and $R$ when necessary to
distinguish which role is associated with the values.
%

\subsection{Preliminaries}
\label{sec:preliminaries}
Private/public key pairs are written
$\langle d_{t,\mathit{id}},\,Q_{t,\mathit{id}}\rangle$,
where $d$ is the private key, $Q$ the public key, $t \in \{e, s\}$ denotes
whether the key is ephemeral or static, and $\mathit{id}$ is the role or
party controlling the key pair.
When irrelevant, we drop the subscript or parts thereof.
Ephemeral key pairs are generated fresh for each instantiation of the protocol
and static key pairs are long-term keys used for authentication.
Static key pairs are suitable for either regular signatures or
challenge-response signatures.
When a party uses regular signatures for authentication, we say that they use
the \emph{signature based authentication method}, or \mSig{} for short.
When a party uses challenge-response signatures for authentication, we say that
they use the \emph{static key authentication method}, or \mStat{} for short.
The latter naming may appear confusing since signature keys are equally static,
but is chosen to make the connection to the \mSpec{} clear.
We adopt the challenge-response terminology for this style of authentication
from~\cite{DBLP:conf/crypto/Krawczyk05}.

\mEdhoc{} relies on \mCose{}~\cite{rfc8152} for elliptic curve operations and
transforming points into bitstrings, and we therefore abstract those as
follows.
Signatures and verification thereof using party $A$'s key pair are
denoted by \mSign{A}$(\cdot)$ and \mVf{A}$(\cdot)$ respectively.
The DH-primitive combining a private key $d$ and a point $P$ is denoted by
$\mDH(d,P)$.
We abuse notation and let these function symbols denote operations on both
points and the corresponding bitstrings.
%

\subsection{Framework Structure}
\label{sec:framework}
\mEdhoc{}'s goal is to establish an \mOscore{} security context,
including session key material \mbox{denoted \mSessKey{}}, and optionally transfer
application data \mADone{}, \mADtwo{} and \mADthree{}.
To accomplish this, the
\mSpec{}~\cite{our-analysis-selander-lake-edhoc-00} gives
a three-message protocol pattern, shown in Figure~\ref{fig:edhocFramework}.
We first describe this pattern and the parts that are common to all key
establishment methods.
Then we describe authentication and derivation of keys in more detail.
The latter is what differentiate the key establishment methods from each other.

\subsubsection{Protocol Pattern}
The first two messages negotiate authentication methods \mMethod{} and a
ciphersuite \mSuites{}.
In \mMethod{}, the initiator $I$ proposes which authentication method each
party should use.
These may differ, leading to four possible combinations:
\mSigSig{}, \mSigStat{}, \mStatSig{} and \mStatStat{}.
We refer to these \emph{combinations} of authentication methods simply as
\emph{methods} to align with the \mSpec{} terminology.
The first authentication method in a combination is the one proposed for the
initiator and the last is the one proposed for the responder.
The responder $R$ may reject the choice of method
or cipher suite with an error message, resulting in negotiation across
multiple \mEdhoc{} sessions.
Our analysis excludes error messages.

\mEdhoc{}'s first two messages also exchange connection identifiers \mCi{} and
\mCr{}, and public ephemeral keys, \mGx{} and \mGy{}.
The connection identifiers \mCi{} and \mCr{}, described in Section 3.1 of the
\mSpec{}, deserve some elaboration.
The \mSpec{} describes these identifiers not as serving a security purpose for
\mEdhoc{}, but only as aiding message routing to the correct \mEdhoc{} processing
entity at a party.
Despite this, the \mSpec{} states that they may be used by \mOscore{}, or other
protocols using the established security context, without restricting how they
are to be used.
Because \mEdhoc{} may need them in clear-text for routing, \mOscore{} cannot
rely on them being secret.
Section 7.1.1 of the \mSpec{} requires the identifiers to be unique.
Uniqueness is defined to mean that $\mCi{} \not = \mCr{}$ for a given session
and the \mSpec{} requires parties to verify that this is the case.
The same section also require that \mOscore{} must be able to retrieve the
security context based on these identifiers.
The intended usage of \mCi{} and \mCr{} by \mOscore{} is not made specific and
therefore it is not clear which properties should be verified.
We verify that the parties agree on the established values.

The two last messages provide identification and authentication.
Parties exchange identifiers for their long-term keys, \mIdcredi{} and \mIdcredr{},
as well as information elements, \mAuthi{} and \mAuthr{}, to authenticate
that the parties control the corresponding long-term keys.
The content of \mAuthi{} and \mAuthr{} depends on the authentication method
associated with the corresponding long-term key.
For example, if $\mMethod{} = \mSigStat{}$, the responder $R$ must either
reject the offer or provide an \mIdcredr{} corresponding to a key pair
$\langle\mPriv{R},\ \mPub{R}\rangle$ suitable for use with challenge-response
signatures, and
compute \mAuthr{} based on the static key authentication method \mStat{}.
In turn, the initiator $I$ must provide an \mIdcredi{} corresponding to a key
pair $\langle\mPriv{I},\ \mPub{I}\rangle$ suitable for a regular signature,
and compute
\mAuthi{} based on the signature based authentication method \mSig{}.

\subsubsection{Authentication}
Regardless of whether \mStat{} or \mSig{} is used to compute \mAuthr{}, a
MAC is first computed over \mIdcredr{}, \mCredr{}, a transcript hash of the
information exchanged so far, and \mADtwo{} if included.
The MAC is the result of encrypting the empty string with the Authenticated
Encryption with Additional Data (AEAD) algorithm from the ciphersuite
\mSuites{}, using the mentioned information as additional data.
The MAC key is derived from the ephemeral key material
\mGx{}, \mGy{}, \mX{} and \mY{}, where $I$
computes $\mDH(\mX,\ \mGy)$ and $R$ computes $\mDH(\mY,\ \mGx)$, both resulting in
the same output in the usual DH way.

\begin{figure}
\centering
\tikzset{>=latex, every msg/.style={draw=thick}, every node/.style={fill=none,text=black}}
\begin{tikzpicture}
    \node (ini) at (0, 0) {Initiator};
    \draw [very thick] (0, -0.25) -- (0,-2.3);
    \draw [very thick] (5.75, -0.25) -- (5.75,-2.3);
    \node (res) at (5.75,0) {Responder};
    \msg{1em}{ini}{res}{\mMsgone: \mMethod, \mSuites, \mGx, \mCi, \mADone};
    \msg{3em}{res}{ini}{\mMsgtwo: \mCi, \mGy, \mCr, \{\mIdcredr, \mAuthr, \mADtwo\}};
    \msg{5em}{ini}{res}{\mMsgthree: \mCr, \{\mIdcredi, \mAuthi, \mADthree\}};
    \draw [line width=1mm] (-0.75,-2.3) -- (0.75,-2.3);
    \draw [line width=1mm] (5.75-0.75,-2.3) -- (5.75+0.75,-2.3);
    \node (padding) at (0,-2.5) {};
    \end{tikzpicture}
    \begin{tabular}{|c|c|c|}
        \hline
        \mMethod & \mAuthi & \mAuthr\\
        \hline
        \mSigSig{} & $\mSign{I}(\cdot)$ & $\mSign{R}(\cdot)$ \\
        \mSigStat{} & $\mSign{I}(\cdot)$ & $\textit{MAC}_R(\cdot)$\\
        \mStatSig{} & $\textit{MAC}_I(\cdot)$ & $\mSign{R}(\cdot)$\\
        \mStatStat{} & $\textit{MAC}_I(\cdot)$ & $\textit{MAC}_R(\cdot)$\\
        \hline
    \end{tabular}
    \caption{Structure of \mEdhoc{}: $\{t\}$ means $t$ is encrypted and integrity
protected.}
\label{fig:edhocFramework}
\end{figure}
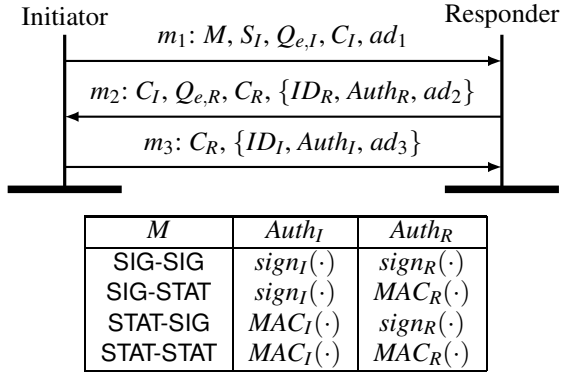

In case $R$ uses the \mSig{} authentication method, \mAuthr{} is $R$'s
signature over the MAC itself and the data that the MAC already covers.
In case $R$ uses the \mStat{} authentication method, \mAuthr{} is simply the
MAC itself.
However, when using \mStat{}, the key for the MAC is derived, not only from the
ephemeral key material, but also from $R$'s long-term key
$\langle\mPriv{R},\ \mPub{R}\rangle$.
For those familiar with \mOptls, this corresponds to the 1-RTT semi-static
pattern computing the MAC key \textsf{sfk} for the \textsf{sfin}
message~\cite{DBLP:conf/eurosp/KrawczykW16}.
The content of \mAuthi{} is computed in the corresponding way for the initiator
$I$.
In Figure~\ref{fig:edhocFramework}, we denote a MAC using a key derived from
both $\langle\mPriv{R},\ \mPub{R}\rangle$ and $\langle\mX,\ \mGx\rangle$ by
$\mathit{MAC}_I$, and a MAC using a key derived from
both $\langle\mPriv{I},\ \mPub{I}\rangle$ and $\langle\mY,\ \mGy\rangle$ by
$\mathit{MAC}_R$.

Parts of the last two messages are encrypted and integrity protected, as
indicated in~\ref{fig:edhocFramework}.
The second message is encrypted by XORing the output of the key derivation
function HKDF (see Section~\ref{sec:keysched}) on to the plain text.
The third message is encrypted and integrity protected by the AEAD algorithm
determined by the ciphersuite \mSuites{}.

\subsubsection{Key Schedule}
\label{sec:keysched}
\begin{figure*}[!h]
\centering
\scalebox{.785}{
\input{kdfdiagram.tex}
}
\caption{Key schedule: Light blue boxes hold DH keys ($P_e, P_I, P_R$),
orange boxes intermediate key material (\mPRKtwo, \mPRKthree, \mPRKfour), and
dark blue boxes keys for \mAead{} or \mXor{} encryption
(\mKtwoe, \mKtwom, \mKthreeae, \mKthreem).
Dashed boxes are conditionals.}
\label{fig:kdfdiagram}
\end{figure*}
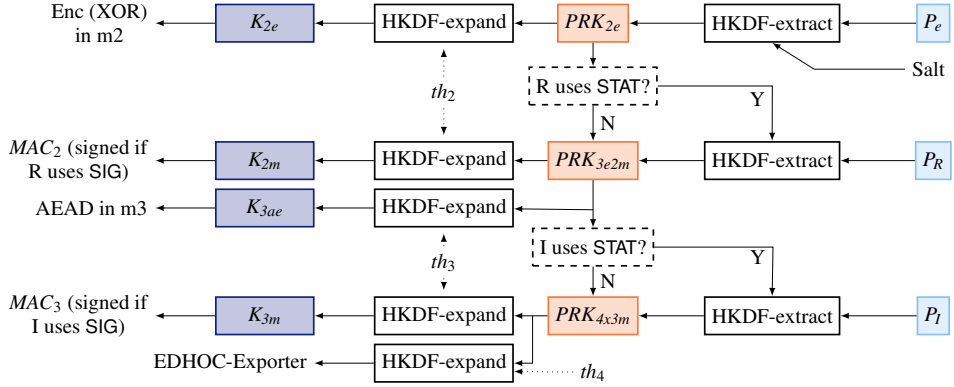

At the heart of \mEdhoc{} is the key schedule depicted in
Figure~\ref{fig:kdfdiagram}.
\mEdhoc{} uses two functions from the \mHkdf{} interface~\cite{rfc5869} to derive keys.
\mHkdfExtract{} 
constructs uniformly distributed key material from random input and a salt,
while \mHkdfExpand{} generates keys from key material and a salt.

The key schedule is rooted in the ephemeral DH key
\mGxy{}, which is computed as $\mDH(\mX, \mGy)$ by $I$ and as $\mDH(\mY, \mGx)$
by $R$.
From \mGxy{}, three intermediate keys \mPRKtwo, \mPRKthree{} and
\mPRKfour{} are derived during the course of protocol execution.
Each of them is used for a specific message in the protocol, and from
these intermediate keys, encryption and integrity keys
(\mKtwoe, \mKtwom{}, \mKthreeae, and \mKthreem) for that message are derived.
The salt for generating \mPRKtwo{} is the empty string.

The protocol uses a running transcript hash $th$, which includes the information
transmitted so far.
The value of the hash, denoted $th_i$ for the $i$th message, is included in key
derivations as shown in Figure~\ref{fig:kdfdiagram}.

Successful protocol execution establishes the session key material \mSessKey{}
for \mOscore{}.
\mSessKey{} can be considered a set that always includes \mGxy{}.
If the initiator uses the \mStat{} authentication method, \mSessKey{} also
includes
$\mDH(\mY,\ \mGi{}) = \mDH(\mPriv{I},\ \mGy)$, which we denote by $P_I$.
If the responder uses the \mStat{} authentication method, it also includes
$\mDH(\mX,\ \mGr{}) = \mDH(\mPriv{R},\ \mGx)$, which we denote by $P_R$.
From the session key material, a key exporter (\mEdhoc-Exporter) based on
\mHkdf{} is used to extract keys required for \mOscore{}.

As an illustrative example of the entire process, we refer to
Figure~\ref{fig:edhocsigstat}, which depicts the protocol pattern, operations
and key derivations for the \mSigStat{} method in more detail.
\begin{figure*}[ht]
\centering
\scalebox{.7}{
\tikzset{>=latex, every msg/.style={draw=thick}, every node/.style={fill=none,text=black}}
\begin{tikzpicture}
    \node (ini) at (0, 0) {Initiator};
    \draw [very thick] (0, -0.5) -- (0,-14.8);
    \draw [very thick] (9, -0.5) -- (9,-14.8);
    \node[below=0.5em of ini,fill=white] {$
    \begin{array}{c}
        \text{Knows}\ \langle\mPriv{I},\ \mPub{I}\rangle,\ \mIdcredi,\ \mIdcredr,\ \mADone,\ \mADthree
    \end{array}
    $};
    \node (res) at (9,0) {Responder};
    \node[below=0.5em of res,fill=white] {$
    \begin{array}{c}
        \text{Knows}\ \langle\mPriv{R},\ \mPub{R}\rangle, \ \mIdcredr,\ \mADtwo
    \end{array}$};
    \action{3.5em}{ini}{Generates $\mMethod,\ \mSuites,\ \mCi,\ \langle\mX{},\ \mGx\rangle$};
    \msg{6.5em}{ini}{res}{\mMsgone: \mMethod, \mSuites, \mGx, \mCi, \mADone};
    \action{7.0em}{res}{$
      \begin{array}{c}
          \text{Generates } \mCr,\ \langle\mY{},\ \mGy\rangle\\
          \ \ P_e = \mDH(\mY,\ \mGx{})\\
          \ \ P_R = \mDH(\mPriv{R},\ \mGx{})\\
        \mTHtwo = \mHash(\mMsgone, \langle \mCi, \mGy, \mCr \rangle)\\
        \mPRKtwo = \mHkdfExtract(\textrm{``\phantom{}''}, P_e) \\
        \mPRKthree = \mHkdfExtract(\mPRKtwo, P_R) \\
        \mKtwom = \mHkdfExpand(\mPRKthree, \mTHtwo) \\
        \mMactwo = \mAead(\mKtwom; \langle \mIdcredr, \mTHtwo, \mCredr, \mADtwo \rangle; \textrm{``\phantom{}''}) \\
        \mKtwoe = \mHkdfExpand(\mPRKtwo, \mTHtwo)
      \end{array}$};
    \msg{21.7em}{res}{ini}{\mMsgtwo: \mCi, \mGy, \mCr, $\overbrace{\mKtwoe\ \mXor\ \langle \mIdcredr, \mMactwo, \mADtwo \rangle}^{\mCipher}$};
    \action{22.5em}{ini}{$
      \begin{array}{c}
        \ P_e = \mDH(\mX,\ \mGy{})\\
        \mPRKtwo = \mHkdfExtract(\textrm{``\phantom{}''}, P_e) \\
        \ \ P_R = \mDH(\mX,\ \mPub{R})\\
        \mPRKfour = \mPRKthree = \mHkdfExtract(\mPRKtwo, P_R) \\
        \mKthreeae = \mHkdfExpand(\mPRKthree, \mTHtwo) \\
        \mTHthree = \mHash(\mTHtwo, \mCipher, \mCr)\\
        \mKthreem = \mHkdfExpand(\mPRKfour, \mTHthree) \\
        \mMacthree = \mAead(\mKthreem; \langle \mIdcredi, \mTHthree, \mCredi, \mADthree \rangle; \textrm{``\phantom{}''}) \\
        \mSigthree = \mSign{I}(\langle \langle \mIdcredi, \mTHthree, \mCredi, \mADthree \rangle, \mMacthree \rangle)
      \end{array}$};
    \msg{35.5em}{ini}{res}{$\mMsgthree: \mCr, \mAead(\mKthreeae; \mTHthree; \langle \mIdcredi, \mSigthree, \mADthree \rangle$)};
    \action{36em}{res}{$
    \begin{array}{c}
        \mTHthree = \mHash(\mTHtwo, \mCipher, \mCr)\\
        \mKthreem = \mHkdfExpand(\mPRKthree, \mTHthree) \\
        \mKthreeae = \mHkdfExpand(\mPRKthree, \mTHthree)
    \end{array}$};
    \draw [line width=1mm] (-2,-14.8) -- (2,-14.8);
    \draw [line width=1mm] (7,-14.8) -- (11,-14.8);
    \end{tikzpicture}
}
    \caption{The \mSigStat{} method.
    Tuples are denoted $\langle\cdot\rangle$, and the hash function \mHash{} is
    as determined by \mSuites{}.}
\label{fig:edhocsigstat}
\end{figure*}

%% file: kdfdiagram.tex

\begin{tikzpicture}[%
    >=latex,              
    start chain=going below,    
    node distance=4mm and 60mm, 
    every join/.style={norm},   
    ]
\tikzset{
terminput/.style={rounded corners},
term/.style={rounded corners},
  base/.style={draw, thick, on chain, on grid, align=center, minimum height=4ex},
  dhkbox/.style={draw=cbsky, fill=cbsky!25, rectangle},
  dhk/.style={base, dhkbox},
  prkbox/.style={draw=cborange, fill=cborange!25, rectangle},
  prk/.style={base, prkbox},
  hkdfext/.style={base, draw=black, fill=none, rectangle},
  hkdfexp/.style={base, draw=black, fill=none, rectangle},
  keybbox/.style={draw=cbnavy, fill=cbnavy!25, rectangle},
  keyb/.style={base, keybbox, text width=4em},
  norm/.style={->, draw, black},
  cond/.style={base, draw=black, dashed, fill=none, rectangle},
  txt/.style={base, draw=none, fill=none}
  }
\node [prk, join] (p2) {\mPRKtwo};
\node [cond, join] (c1) {R uses \mStat?};

\node [prk, below=6mm of c1.south] (p3) {\mPRKthree};
\draw [->, norm] (c1.south) -- (p3.north) node[midway, right] {N};

\node [cond, join, below=8mm of p3.south] (c2) {I uses \mStat?};
\node [prk, below=5mm of c2.south] (p4) {\mPRKfour};
\draw [->, norm] (c2.south) -- (p4.north) node[midway, right] {N};

\node [hkdfext, right=3cm of p3] (h3) {\mHkdfExtract};
\node [hkdfext, right=3cm of p4] (h5) {\mHkdfExtract};

\node [hkdfexp, shape border rotate=180, left= 2.5cm of p4] (h6) {\mHkdfExpand};
\node [keyb, join, left=3cm of h6] (k3) {\mKthreem};
\node [hkdfexp, shape border rotate=180, below= 0.8cm of h6] (h9) {\mHkdfExpand};
\node [txt, join, left=1cm of h9.west] (t4) {EDHOC-Exporter};

\node [hkdfexp, shape border rotate=180, left= 2.5cm of p3] (h4) {\mHkdfExpand};
\node [keyb, join, left=3cm of h4] (k2) {\mKtwom};

\node [hkdfexp, shape border rotate=180, left= 2.5cm of p2] (h2) {\mHkdfExpand};
\node [keyb, join, left=3cm of h2] (k1) {\mKtwoe};

\node [hkdfexp, shape border rotate=180, below= 0.8cm of h4] (h8) {\mHkdfExpand};
\node [keyb, below=0.8cm of k2] (k2b) {\mKthreeae};

\node [txt, left=1cm of k1.west] (t1) {Enc (XOR) \\ in m2};
\node [txt, left=1cm of k2.west] (t2) {\mMactwo~(signed if \\ R uses \mSig)};
\node [txt, left=1cm of k2b.west] (t2b) {\mAead\ in m3};
\node [txt, left=1cm of k3.west] (t3) {\mMacthree~(signed if \\ I uses \mSig)};

\draw [->, norm] (k1.west) -- (t1.east);
\draw [->, norm] (k2.west) -- (t2.east);
\draw [->, norm] (k2b.west) -- (t2b.east);
\draw [->, norm] (k3.west) -- (t3.east);

\draw [->, norm] (p3.south) ++(0,-0.5) -- (h8);
\draw [->, norm] (h8) -- (k2b);
\draw [->, norm] (p2) -- (h2); 
\draw [->, norm] (c1.east) -- ++(1.92,0) -- (h3.north) node[midway,above left] {Y};
\draw [->, norm] (h3.west) -- (p3.east);
\draw [->, norm] (p3) -- (h4); 
\draw [->, norm] (c2.east) -- ++(1.99,0) -- (h5.north) node[midway,above left] {Y};
\draw [->, norm] (h5.west) -- (p4.east);
\draw [->, norm] (p4) -- (h6);
\draw [->, norm] (p4.west) ++(-0.25,-0) -- ++(0,-0.8) -- (h9.east);

\node [hkdfext, right=3cm of p2] (h1) {\mHkdfExtract};
\node [dhk, right=2.7cm of h1] (p0) {$\mGxy$};
\node [terminput, text width=2em, below = 0.2cm of p0] (u1) {Salt};
\draw [->] (h1.west) -- (p2.east);
\draw [->] (u1.west) -- ++(-1.52,0) -- (h1.south);
\draw [->] (p0.west) -- (h1.east);

\node [dhk, right = 2.7cm of h3] (u2) {$P_R$};
\draw [->, norm] (u2.west) -- (h3.east);

\node [dhk, right = 2.7cm of h5] (u3) {$P_I$};
\draw [->, norm] (u3.west) -- (h5.east);

\node [term, above = 0.55cm of h4] (u5) {\mTHtwo};
\draw [->, dotted, shorten >=1mm] (u5) -- (h4);
\draw [->, dotted, shorten >=1mm] (u5) -- (h2);

\node [term, above = 0.3cm of h6] (u6) {\mTHthree};
\draw [->, dotted, shorten >=1mm] (u6) -- (h6);
\draw [->, dotted, shorten >=1mm] (u6) -- (h8);

\node [term, below=0.35cm of p4] (u7) {\mTHfour};
\draw [-> , dotted ] (u7.west) -- ([yshift=-0.4em] h9.east);


%
%
\end{tikzpicture}

%% file: edhocFormalization.tex
The \mEdhoc{} \mSpec{} \cite{our-analysis-selander-lake-edhoc-00} claims
that \mEdhoc{} satisfies many security properties, but these are imprecisely
expressed and motivated.
In particular, there is no coherent adversary model.
It is therefore not clear in which context properties should be verified.
We resolve this by clearly specifying an adversary model, in which we can verify
properties.

\subsection{Adversary Model}\label{sec:threat-model}
We verify \mEdhoc{} in the symbolic Dolev-Yao model, with idealized
cryptographic primitives, e.g, encrypted messages can only be
decrypted using the key, no hash collisions exist etc.
The adversary controls the
communication channel, and can interact with an unbounded number of sessions
of the protocol, dropping, injecting and modifying messages to their liking.

In addition to the basic Dolev-Yao model, we also consider two more adversary
capabilities, namely long-term key reveal and ephemeral key reveal.
Long-term key reveal models the adversary compromising a party $A$'s
long-term private key \mPriv{A} at time $t$, and we denote this event by
$\mRevLTK^t(A)$.
The event $\mRevEph^t(A, k)$ represents that the adversary learns
the ephemeral private key \mPrivE{A} used by party $A$ at time $t$ in a session
establishing key material $k$.
These two capabilities model the possibility to store and operate on
long-term keys in a secure module, whereas ephemeral keys
may be stored in a less secure part of a device.
This is more granular and realistic than assuming that the adversary has equal
opportunity to access both types of keys.

We now define and formalize the security properties we are interested in, and
then describe how we encode them into \mTamarin{}.
The adversary model becomes part of the security properties themselves.

\subsection{Formalization of Properties}
\label{sec:desired-properties}
We use the \mTamarin{} verification
tool~\cite{DBLP:conf/cav/MeierSCB13} to encode the model and verify properties.
This tool uses a fragment of temporal first order logic to reason about
events and knowledge of the parties and of the adversary.
For conciseness we use a slightly different syntax than
that used by \mTamarin{}, but which has a direct mapping to \mTamarin{}'s logic.

Event types are predicates over global states of system execution.
Let $E$ be an event type and let $t$ be a timestamp associated with a point in a
trace.
Then $E^{t}(p_i)_{i\in\mathbb{N}}$ denotes an event of type $E$ associated with
a sequence of parameters $(p_i)_{i\in\mathbb{N}}$ at time $t$ in that trace.
In general, more than one event may have the same timestamp and hence
timestamps form a quasi order, which we denote by $t_1 \lessdot t_2$ when $t_1$
is before $t_2$ in a trace.
We define $\doteq$ analogously.
However, two events of the same type cannot have the same timestamp, so
$t_1 \doteq t_2$ implies $E^{t_1} = E^{t_2}$.
Two events having the same timestamp does not imply the there is a fork in the
trace, only that the two events happen simultaneously.
This notation corresponds to \mTamarin{}'s use of action facts
$E(p_i)_{i\in\mathbb{N}}@t$.

The event $\mK^t(p)$ denotes that the adversary knows a parameter $p$ at
time $t$.
Parameters are terms in a term algebra of protocol specific operations and
generic operations, e.g., tuples $\langle\cdot\rangle$.
Intuitively, $\mK^t(p)$ evaluates to true when $p$ is in
the closure of the
parameters the adversary observed by interacting with parties using the
protocol, under the Dolev-Yao message deduction operations and
the advanced adversary capabilities up until time $t$.
For a more precise definition of knowledge management, we refer to~\cite{DBLP:conf/cav/MeierSCB13}.
An example of a formula is
\[
    \forall t, k, k'\mLogicDot \mK^{t}(\langle k, k'\rangle)\ \rightarrow\ 
\mK^{t}(k) \land \mK^{t}(k'),
\]
expressing that if there is a time $t$ when the adversary knows the tuple
$\langle k, k'\rangle$, then the adversary knows both
$k$ and $k'$ at the same point in time.

An initiator $I$ considers the
protocol run started when it sends a message \mMsgone{} (event type \mIStart)
and the run completed after sending a message \mMsgthree{} (event type
\mIComplete).
Similarly, a responder $R$ considers the run started upon receiving
a \mMsgone{} (event type \mRStart), and completed upon receiving a \mMsgthree{}
(event type \mRComplete).
%

\subsubsection{Perfect Forward Secrecy (PFS)}
\label{sec:secrecy}
Informally, PFS captures the idea that session key material remains secret
even if a long-term key leaks in the future.
We define PFS for session key material \mSessKey{} as \mPredPfs{} in
Figure~\ref{fig:props}.

\begin{figure*}
\begin{align*}
    \mPredPfs \triangleq\ & \forall I, R, \mSessKey, t_2, t_3\mLogicDot
    \mK^{t_3}(\mSessKey)\  \land\ 
    (\mIComplete^{t_2}(I, R, \mSessKey)\, \lor\, \mRComplete^{t_2}(I, R, \mSessKey))
    \rightarrow\\
    &(\exists t_1\mLogicDot \mRevLTK^{t_1}(I) \land t_1 \lessdot t_2)
    \ \lor\ (\exists t_1\mLogicDot \mRevLTK^{t_1}(R) \land t_1 \lessdot t_2)
    \ \lor\ (\exists t_1\mLogicDot \mRevEph^{t_1}(R, \mSessKey))
    \ \lor\ (\exists t_1\mLogicDot \mRevEph^{t_1}(I, \mSessKey))
\end{align*}
\begin{align*}
    \mPredInjI \triangleq\ &
    \forall I, R, \mSessKey, S, t_2\mLogicDot \mIComplete^{t_2}(I, R, \mSessKey, S)
    \rightarrow\\
    &(\exists t_1\mLogicDot \mRStart^{t_1}(R, \mSessKey, S) \land t_1 \lessdot t_2)
    \land (\forall I' R' t_1' \mLogicDot \mIComplete^{t_1'}(I' , R', \mSessKey, S)
        \rightarrow t_1' \doteq t_1)
    \ \ \lor\ \ (\exists t_1\mLogicDot \mRevLTK^{t_1}(R) \land t_1 \lessdot t_2)
\end{align*}
\begin{align*}
    \mPredInjR \triangleq\ &
    \forall I, R, \mSessKey, S, t_2\mLogicDot \mRComplete^{t_2}(I, R, \mSessKey, S)
    \rightarrow\\
    &(\exists t_1\mLogicDot \mIStart^{t_1}(I, R, \mSessKey, S) \land t_1 \lessdot t_2)
    \land (\forall I' R' t_1' \mLogicDot \mRComplete^{t_1'}(I' , R', \mSessKey, S)
        \rightarrow t_1' \doteq t_1)
    \ \ \lor\ \ (\exists t_1\mLogicDot \mRevLTK^{t_1}(I) \land t_1 \lessdot t_2).
\end{align*}
\begin{align*}
    \mPredImpI \triangleq\ &
    \forall I, R, \mSessKey, S, t_1\mLogicDot \mIComplete^{t_1}(I, R, \mSessKey, S)
    \rightarrow\\
      &(\forall I', R', S', t_2\mLogicDot \mRComplete^{t_2}(I', R', \mSessKey, S') \rightarrow
             (I=I' \land R=R' \land S=S'))\\
      &\land (\forall I', R', S', t_1'\mLogicDot
        (\mIComplete^{t_1'}(I', R', \mSessKey, S') \rightarrow t_1' \doteq t_1
        )\\
      &\ \ \ \lor(\exists t_0\mLogicDot \mRevLTK^{t_0}(R) \land t_0 \lessdot t_1)
    \lor(\exists t_0\mLogicDot \mRevEph^{t_0}(R, \mSessKey))
    \lor(\exists t_0\mLogicDot \mRevEph^{t_0}(I, \mSessKey)).
\end{align*}
\caption{Formalization of security properties and adversary model.}
\label{fig:props}
\end{figure*}

The first parameter $I$ of the \mIComplete{} event represents the
initiator's identity,
and the second, $R$, represents that $I$ believes $R$ to be playing
the responder role.
The third parameter, \mSessKey{}, is the established session key material.
The parameters of the \mRComplete{} event are defined analogously.
Specifically, the first parameter of \mRComplete{} represents the identity of
whom $R$ believes is playing the initiator role.
The essence of the definition is that an adversary only knows \mSessKey{} if they
compromised one of the
parties long-term keys before that party completed the run, or if the adversary
compromised any of the ephemeral keys at any time after a party starts
its protocol run.
One way the definition slightly differs from the corresponding \mTamarin{} lemma
is that \mTamarin{} does not allow a disjunction on the left-hand side of an
implication in a universally quantified formula.
In the lemma, therefore, instead of the disjunction
$\mIComplete^{t_2}(I, R, \mSessKey)\, \lor\,  \mRComplete^{t_2}(I, R, \mSessKey)$,
we use a single action parametrized by $I$, $R$, and \mSessKey{} to signify that
\emph{either} party has completed their role.
%

\subsubsection{Authentication}
\label{sec:authenticationDef}
We prove two different flavors of authentication, the first being classical
\emph{injective agreement} following Lowe~\cite{DBLP:conf/csfw/Lowe97a}, and
the second being an implicit agreement property.
Informally, injective agreement guarantees to an initiator $I$ that whenever
$I$ completes a run ostensibly with a responder $R$,
then $R$ has been engaged in the protocol as a responder,
and this run of $I$ corresponds to a unique run of $R$.
In addition, the property guarantees to $I$ that the two parties agree on a set
$S$ of parameters associated with the run, including, in particular, the
session key material \mSessKey{}.
However, we will treat \mSessKey{} separately for clarity.
On completion, $I$ knows that $R$ has access to the session key material.
The corresponding property for $R$ is analogous.

Traditionally, the event types used to describe injective agreement are called
\emph{Running} and \emph{Commit}, but to harmonize the presentations of
authentication and PFS in this section, we refer to these event types as
\mIStart{} and \mIComplete{} respectively for the initiator, and
\mRStart{} and \mRComplete{} for the responder.
For the initiator role we define injective agreement as given by
\mPredInjI{} in Figure~\ref{fig:props}.

The property captures that for an initiator $I$, either the injective agreement
property as described above holds, or the long-term key of the believed
responder $R$ has been compromised before $I$ completed its role.
Had the adversary compromised $R$'s long-term key, they could have generated a
message of their liking (different from what $R$ agreed on) and signed this or
computed a $\mathit{MAC}_R$ based on \mPubE{I}, \mPriv{R} and their own chosen
ephemeral key pair $\langle\mPrivE{R},\ \mPubE{R}\rangle$.
This places no restrictions on the ephemeral
key reveals, or on the reveal of the initiator's long-term key.
For the responder we define the property \mPredInjR{} as in
Figure~\ref{fig:props}.
%

Unlike PFS, not all \mEdhoc{} methods enjoy the injective agreement property.
Hence, we show for all methods a form of \emph{implicit agreement} on all the
parameters mentioned above.
We take inspiration from the computational model definitions of implicit
authentication, proposed by~\cite{DBLP:conf/csfw/GuilhemFW20}, to
modify classical injective agreement into an implicit property.
A small but important difference between our definition and theirs, is that
they focus on
authenticating a key and related identities, whereas we extend the more general
concept of agreeing on a set of parameters, starting from the idea of injective
agreement~\cite{DBLP:conf/csfw/Lowe97a}.
We use the term \emph{implicit} in this context to denote that a party $A$
assumes that any other party $B$ who knows the session key material \mSessKey{} must
be the intended party, and that $B$ (if honest) will also agree on a set
$S$ of parameters computed by the protocol, one of which is \mSessKey{}.
When implicit agreement holds for both roles, upon completion, $A$ is guaranteed
that $A$ has been or is engaged in exactly one protocol run with $B$ in the
opposite role, and that $B$ has been or will be able to agree on $S$.
The main difference from injective agreement is that $A$ concludes that if
$A$ sends the last message and this reaches $B$, then $A$ and $B$ have agreed
on $I$, $R$ and $S$.
While almost full explicit key authentication, as defined
by~\cite{DBLP:conf/csfw/GuilhemFW20}, is a
similar property, our
definition does not require key confirmation, so our definition is closer to
their definition of implicit authentication.
In the \mTamarin{} model we split the property into one lemma for
$I$ (\mPredImpI{}) and one for $R$ (\mPredImpR{}) to save memory during
verification.
We show only the definition for $I$ in Figure~\ref{fig:props}, because it is
symmetric to the one for $R$.

For implicit agreement to hold for the initiator $I$, the ephemeral keys
can never be revealed.
Intuitively, the reason for this is that the implicit agreement relies on that
whomever knows the session key material is the intended responder.
An adversary with access to the ephemeral keys and the public keys of
both parties can compute the session key material produced by all methods.
However, the responder $R$'s long-term key can be revealed after $I$ completes
its run, because the adversary is still unable to compute $P_e$.
The initiator's long-term key can also be revealed at any time without affecting
$I$'s guarantee for the same reason.
%

\subsubsection{Agreed Parameters}
\label{sec:agreedParams}
The initiator $I$ gets injective and implicit agreement guarantees on the
following partial set $S_P$ of parameters:
\begin{itemize}
    \item the roles played by itself and its peer,
    \item responder identity,
    \item session key material (which varies depending on \mEdhoc{} method),
    \item context identifiers \mCi{} and \mCr{}, and
    \item cipher suites \mSuites{}.
\end{itemize}
Because \mEdhoc{} aims to provide identity protection for $I$, there is no
injective agreement guarantee for $I$ that $R$ agrees on the initiator's
identity.
For the same reason, there is no such guarantee for $I$ with respect to
the $P_I$ part of the session key material when $I$ uses the \mStat{}
authentication method.
There is, however, an implicit agreement guarantee for $I$ that $R$ agrees on
$I$'s identity and the full session key material.
Since $R$ completes after $I$, $R$ can get injective agreement guarantees on
more parameters, namely also the initiator's identity and the full session key
material for all methods.
The full set of agreed parameters $S_F$ is $S_P \cup \{I, P_I\}$
when $P_I$ is
part of the session key material, and $S_P \cup \{I\}$ otherwise.
%

\subsubsection{Inferred Properties}
From the above, other properties can be inferred to hold in our adversary model.
Protocols where a party does not get confirmation that their peer knows the
session key material may be susceptible to
\emph{Key-Compromise Impersonation (KCI)}
attacks~\cite{DBLP:conf/ima/Blake-WilsonJM97}.
Attacks in this class allow an adversary in possession of a party $A$'s secret
long-term key to coerce $A$ to complete a
protocol run believing it authenticated a certain peer $B$, but where $B$ did
not engage with $A$ at all in a run.
Because both our above notions of agreement ensure agreement on identities,
roles and session key material, all methods passing verification of those are
also resistant to KCI attacks.

If a party $A$ can be coerced into believing it completed a run with $B$, but
where the session key material is actually shared with $C$ instead, the 
protocol is vulnerable to an \emph{Unknown Key-Share (UKS)}
attack~\cite{DBLP:conf/ima/Blake-WilsonJM97}.
For the same reason as for KCI, any method for which our agreement
properties hold is also resistant to UKS attacks.

From the injective agreement properties it follows that each party is assured
the identity of its peer upon completion.
Therefore, the agreement properties also capture \emph{entity authentication}.
%

\subsection{\mTamarin{}}
\label{sec:tamarin}
We chose \mTamarin{} to model and verify \mEdhoc{} in the symbolic model.
It is an interactive verification tool in which models are specified
as multi-set rewrite rules that define a transition relation.
The elements of the multi-sets are facts representing the global system
state.
Rules are equipped with event annotations called actions.
Sequences of actions make up execution traces, over which
logic formulas are defined.

Multi-set rewrite rules are written $ l \ifarrow[e] r $,
where $l$ and $r$ are multi-sets of facts, and $e$ is a multi-set of actions.
Facts and actions are $n$-ary predicates over a term algebra, which defines a
set of function symbols, variables and names.
\mTamarin{} checks equality of these terms under an equational theory $E$.
For example, one can write $ dec(enc(x,y),y) =_E x $
to denote that symmetric decryption reverses the encryption operation under $E$.
The equational theory $E$ is fixed per model, and hence we omit the subscript.
\mTamarin{} supports let-bindings and tuples as syntactic sugar to simplify
model definitions.
It also provides built-in rules for Dolev-Yao adversaries and for
managing their knowledge.
We implement events using actions, and parameters associated with events using
terms of the algebra.

\subsubsection{Protocol Rules and Equations}
\mTamarin{} allows users to define new function symbols, equational theories
and rules, which are added to the set of considered rules during verification.
For example, in our model we have a symbol to denote authenticated encryption,
for which \mTamarin{} produces a rule of the following form:
\begin{small}
\begin{verbatim}
[!KU(k), !KU(m), !KU(ad), !KU(ai)] --[]->
    [!KU(aeadEncrypt(k, m, ad, ai))]
\end{verbatim}
\end{small}
to denote that if the adversary knows a key \mT{k}, a message \mT{m}, the
authenticated data \mT{ad}, and an algorithm \mT{ai}, then they can construct
the encryption, and thus get to know the message
\mT{aeadEncrypt(k, m, ad, ai)}.
%

\subsection{\mTamarin{} Encoding of \mEdhoc{}}
\label{sec:modeling}
We model all four methods of \mEdhoc{}, namely
\mSigSig, \mSigStat, \mStatSig{} and \mStatStat.
Because the methods share a lot of common structure, we derive
their \mTamarin-models from a single specification written with the aid of the
M4 macro language.
To keep the presentation brief, we only present the \mStatSig{} metohod, as it
illustrates the use of two different asymmetric authentication methods
simultaneously.
The full \mTamarin{} code for all models can be found at~\cite{edhocTamarinRepo}.
Variable names used in the code excerpts here are sometimes shortened compared
to the model itself to fit the paper format.

\subsubsection{Primitive Operations}
Our model uses the built-in theories of exclusive-or and DH operations, as
in~\cite{DBLP:conf/csfw/DreierHRS18,DBLP:conf/csfw/SchmidtMCB12}.
Hashing is modeled via the built-in hashing function symbol augmented
with a public constant as additional input, modelling different
hash functions.
The HKDF interface is represented by \mT{expa} for the
expansion operation and \mT{extr} for the extraction operation.
Signatures use \mTamarin's built-in theory for \mT{sign} and \mT{verify}
operations.
For \mAead{} operations on key \mT{k}, message \mbox{\mT{m}}, additional data \mT{ad}
and algorithm identifier \mT{ai}, we use \mT{aeadEncrypt(m, k, ad, ai)}
for encryption.
Decryption with verification of the integrity is defined via the equation
\begin{small}\begin{verbatim}
aeadDecrypt(aeadEncrypt(m, k, ad, ai),
    k, ad, ai) = m.
\end{verbatim}\end{small}
The integrity protection of AEAD covers \mT{ad}, and this equation hence requires
an adversary to know \mT{ad} even if they only wish to decrypt the data.
To enable the adversary to decrypt without needing to verify the integrity
we add the equation
\begin{small}\begin{verbatim}
decrypt(aeadEncrypt(m, k, ad, ai), k, ai) = m.
\end{verbatim}\end{small}
The latter equation is not used by honest parties.

\subsubsection{Protocol Environment and Adversary Model}
We model the binding between a party's identity and their long-term
key pairs using rules for \mSig{}- and \mStat{}-based methods separately.
\begin{small}
\begin{verbatim}
rule registerLTK_SIG:
    [Fr(~ltk)] --[UniqLTK($A, ~ltk)]->
        [!LTK_SIG($A, ~ltk),
         !PK_SIG($A, pk(~ltk)),
         Out(<$A, pk(~ltk)>)]
rule registerLTK_STAT:
    [Fr(~ltk)] --[UniqLTK($A, ~ltk)]->
        [!LTK_STAT($A, ~ltk),
         !PK_STAT($A, 'g'^~ltk),
         Out(<$A, 'g'^~ltk>)]
\end{verbatim}
\end{small}
The fact \verb|Fr(~ltk)| creates a fresh term \mT{ltk}, representing a long-term
secret key, not known to the adversary.
The fact \verb|Out(<$A, pk(~ltk)>)| sends the identity of the party
owning the long-term key and the corresponding public key to the adversary.
The event \mT{UniqLTK} together with a corresponding restriction models the fact
that the each party is associated with exactly one long-term key.
Consequently, an adversary cannot register additional long-term keys for an
identity.
In line with the \mEdhoc{} \mSpec{}, this models an external mechanism
ensuring that long term keys are bound to correct identities, e.g.,
a certificate authority.

We rely on \mTamarin's{} built-in message deduction rules for a Dolev-Yao adversary.
To model an adversary compromising long-term keys, i.e., events of type
\mRevLTK{}, and revealing ephemeral keys, i.e., events of type
\mRevEph{}, we use standard reveal rules.
The timing of reveals as modelled by these events is important.
The long-term keys can be revealed on registration, before protocol execution.
The ephemeral key of a party can be revealed when the party completes,
i.e., at events of type \mIComplete{} and \mRComplete.~\footnote{A stronger, and perhaps more realistic, model would reveal ephemeral keys upon
creation at the start of the run, but we failed to get \mTamarin{} to
terminate on this.}

\subsubsection{Protocol Roles}
We model each method of the protocol with four rules: \mT{I1}, \mT{R2}, \mT{I3}
and \mT{R4} (with the method suffixed to the rule name).
Each of these represent one step of the protocol as run by the initiator $I$
or the responder $R$.
The rules correspond to the event types \mIStart, \mRStart, \mIComplete,  and
\mRComplete, respectively.
Facts prefixed with \mT{StI} carry state information between \mT{I1} and \mT{I3}.
A term unique to the current thread, \mT{tid}, links two rules to a given state fact.
Similarly, facts prefixed with \mT{StR} carry state information between the
responder role's rules.
Line 28 in the \mT{R2\_STAT\_SIG} rule shown below illustrate one such use of state
facts.

We do not model the error message that $R$ can send in response to message
\mMsgone, and hence our model does not
capture the possibility for $R$ to reject $I$'s offer.

We model the XOR encryption of \mT{CIPHERTEXT\_2} with the key \mT{K\_2e} using
\mTamarin{}'s built in theory for XOR, and allow each term of the encrypted
element to be attacked individually.
That is, we first expand \mT{K\_2e} to as many key-stream terms as there are
terms in the plaintext tuple by applying the \mHkdfExpand{} function to unique
inputs per term.
We then XOR each term in the plaintext with its own key-stream term.
This models the \mSpec{} closer than if we would have XORed \mT{K\_2e} as a
single term onto the plaintext tuple.
The XOR encryption can be seen in lines 19--22 in the listing of
\mT{R2\_STAT\_SIG} below.
\begin{small}
\begin{verbatim}
1 rule R2_STAT_SIG:
2 let
3    agreed = <CS0, CI, ~CR>
4    gx = 'g'^xx
5    data_2 = <'g'^~yy, CI, ~CR>
6    m1 = <'STAT', 'SIG', CS0, CI, gx>
7    TH_2 = h(<$H0, m1, data_2>)
8    prk_2e = extr('e', gx^~yy)
9    prk_3e2m = prk_2e
10   K_2m = expa(<$cAEAD0, TH_2, 'K_2m'>,
11               prk_3e2m)
12   protected2 = $V // ID_CRED_V
13   CRED_V = pkV
14   extAad2 = <TH_2, CRED_V>
15   assocData2 = <protected2, extAad2>
16   MAC_2 = aead('e', K_2m, assocData2,
                  $cAEAD0)
17   authV = sign(<assocData2, MAC_2>, ~ltk)
18   plainText2 = <$V, authV>
19   K_2e = expa(<$cAEAD0, TH_2,
                 'K_2e'>, prk_2e)
20   K_2e_1 = expa(<$cAEAD0, TH_2,
                   'K_2e', '1'>, prk_2e)
21   K_2e_2 = expa(<$cAEAD0, TH_2,
                   'K_2e', '2'>, prk_2e)
22   CIPHERTEXT_2 = <$V XOR K_2e_1,
                     authV XOR K_2e_2>
23   m2 = <data_2, CIPHERTEXT_2>
24   exp_sk = <gx^~yy>
25 in
26   [!LTK_SIG($V, ~ltk), !PK_SIG($V, pkV),
      In(m1), Fr(~CR), Fr(~yy), Fr(~tid)]
27   --[ExpRunningR(~tid, $V, exp_sk, agreed),
        R2(~tid, $V, m1, m2)]->
28   [StR2_STAT_SIG($V, ~ltk, ~yy, prk_3e2m,
              TH_2, CIPHERTEXT_2, gx^~yy,
              ~tid, m1, m2, agreed),
29    Out(m2)]
\end{verbatim}
\end{small}

To implement events and
to bind them to parameters, we use actions.
For example, the action \verb|ExpRunningR(~tid, $V, exp_sk, agreed)| in line 27
above implements binding of an event of type \mRStart{} to the parameters and session key
material.

As explained in Section~\ref{sec:agreedParams}, it is not possible to show
injective agreement on session key material when it includes
$P_I$ (not visible in the rule \mT{R2\_STAT\_SIG}).
Therefore, we use certain actions to implement events that include $P_I$ in the
session key material and other actions that do not.
Session key material which includes (resp. does not include) $P_I$ is referred
to as \mT{imp\_sk} (resp. \mT{exp\_sk}) in the
\mTamarin{} model.
In the case of \mSigSig{} and \mSigStat, therefore, \mT{imp\_sk} is the same as
\mT{exp\_sk}.
%

\subsection{\mTamarin{} Encoding of Properties}
\label{sec:propertyFormalization}
The properties and adversary model we defined in
Section~\ref{sec:desired-properties} translate directly into \mTamarin's logic,
using the straightforward mapping of events to the actions emitted from the model.
As an example, we show the lemma for verifying the property \mPredPfs.
\begin{small}
\begin{verbatim}
1 lemma secrecyPFS:
2 all-traces
3  "All u v sk #t3 #t2.
4   (K(sk)@t3 & CompletedRun(u, v, sk)@t2) ==>
5       ( (Ex #t1. LTKRev(u)@t1 & #t1 < #t2)
6       | (Ex #t1. LTKRev(v)@t1 & #t1 < #t2)
7       | (Ex #t1. EphKeyRev(sk)@t1))"
\end{verbatim}
\end{small}
The action \mT{CompletedRun(u, v, sk)} in line 4 is
emitted by both the rules \mT{I3} and \mT{R4}, and corresponds
to the disjunction of events $\mIComplete^{t_2} \lor \mRComplete^{t_2}$ in the
definition of \mPredPfs{} in Section~\ref{sec:secrecy}.
Similarly, \mT{EphKeyRev(sk)} in line 7 models that the ephemeral
key is revealed for either $I$ or $R$, or both.
%